\documentclass[12pt]{article}
\usepackage{epsfig} 
\usepackage{graphicx}  
\usepackage{axodraw}
\usepackage[includeall, letterpaper, body={7in, 9in}, left=.75in, top=1in]{geometry}

\newcommand{\be}{\begin{equation}}  
\newcommand{\ee}{\end{equation}}  
\newcommand{\bea}{\begin{eqnarray}}     
\newcommand{\eea}{\end{eqnarray}}

\begin{document}
\setcounter{page}{0}
\thispagestyle{empty} 
\begin{flushright}
ANL-HEP-PR-06-47
\end{flushright}

\vspace{1cm}

\begin{center}

{\Large \bf Measuring the $W$-$t$-$b$ Interaction at the ILC}\\[1 cm]
{\Large
Puneet Batra and Tim M.P. Tait\\[.5 cm]}

{\small {\it
 High Energy Physics Division, Argonne National Lab, Argonne, IL 60439, USA}}
\end{center}

\vspace{.8cm}
\vspace{0.3cm} {\small \center \noindent \textbf{Abstract} \\[0.3cm]
\noindent } 

The large top quark mass suggests that the top plays a pivotal role in
Electroweak symmetry-breaking dynamics and, as a result, may have
modified couplings to Electroweak bosons. Hadron colliders can provide
measurements of these couplings at the $\sim 10\%$ level, and one of
the early expected triumphs of the International Linear Collider is to
reduce these uncertainties to the per cent level.  In this article, we
propose the first direct measurement of the Standard Model $W$-$t$-$b$
coupling at the ILC, from measurements of $t \bar{t}$-like signals
{\em below} the $t \bar{t}$ production threshold. We estimate that the
ILC with $100~{\rm fb}^{-1}$ can measure a combination of the coupling
and top width to high precision, and when combined with a direct
measurement of the top width from the above-threshold scan, results in
a model-independent measurement of the $W$-$t$-$b$ interaction of the
order of $\sim 3 \%$.  \vfill 
\newpage
\pagestyle{plain}
\section{Introduction}
\label{sec:intro}

The mystery of electroweak symmetry breaking (EWSB)
is the foremost problem in particle physics.  Near-future colliders such as 
the Large Hadron Collider (LHC) and the International Linear Collider (ILC) 
are primarily motivated by their unique ability to make measurements 
which will reveal the mechanism of electroweak symmetry breaking.  In the very
least, a light Higgs is expected to be discovered.  However, it is generally
accepted that the physics at TeV energies will prove to be much richer than
the minimal standard model (SM), and that the new paradigm will tie 
together many interesting puzzles, including the large hierarchy between the 
Planck and the electroweak scales, the nature of dark matter 
and dark energy, neutrino masses, and the puzzling pattern of flavor in 
both the quark and lepton sectors of the Standard Model.  

At the center of many of these mysteries is the top quark.  Its large
mass, itself a manifestation of the broken electroweak symmetry,
indicates that whatever the nature of the symmetry-breaking, it was
communicated strongly to the top.  In the minimal Standard model, this
is represented by a large top Yukawa interaction and results in a
special role for the top in Higgs physics.  In the SM, the
interactions with the gauge and Higgs bosons are predicted by the
gauge structure, top mass, and unitarity of the CKM matrix.  In models
of physics beyond the Standard Model, the top often plays a special
role in driving the electroweak symmetry breaking dynamics.  In order
to explore the possibility that the top plays a special role in EWSB,
precise measurements of all top couplings are needed.  In this
article, we will explore how a future ILC can add to our knowledge of
the $W$-$t$-$b$ interaction strength by looking for off-shell top
quarks produced when the collider is running below the $t\bar{t}$
production threshold.

The charged current interaction of the top with the bottom quark and $W$ boson 
is predicted by the SM to be purely left-chiral with interaction strength,
\bea
g_{Wtb} & \sim & g V_{tb} \sim g ,
\eea
where $g = e / \sin \theta_W$ is the $SU(2)_W$ coupling of the SM and
$V_{tb}$ can be inferred from other elements using unitarity to be 
$0.9990-0.9992$ at the $90\%$ CL\cite{Eidelman:2004wy}.  
This coupling is of great practical importance in top
physics, because it leads to the predominant decay mode $t \rightarrow W b$.
The observation of top in its expected channel (and lack of observation of
other channels) \cite{Erbacher:2005ig}
is an indication that this decay
is much larger than any competing decay mode.  However, it does not allow for
a measurement of the interaction, because the branching ratio ($BR$) is
essentially unity, and thus does not depend sensitively on $g_{Wtb}$.

In models
of physics beyond the Standard Model, the interaction may deviate from its
SM expectation due to new strong dynamics \cite{Carlson:1994bg,Agashe:2005vg}, 
mixing of the top
(and/or bottom) with a new vector-like family of quarks 
\cite{Dobrescu:1997nm,Arkani-Hamed:2002qy}, or through mixing of 
the $W$ boson with some heavier $W^\prime$
that prefers to couple to top \cite{Chivukula:1995gu}.  
More generically, we can place bounds on the
gauge invariant operators which extend the SM, and use these bounds to obtain
information about the parameters of {\em any} theory which predicts a
modification of the $W$-$t$-$b$ interaction.  If we choose to realize the
$SU(2)_L \times U(1)_Y$ of the SM non-linearly \cite{Peccei:1989kr}, 
as may be appropriate for
theories in which EWSB is through unspecified or incalculable strong dynamics,
the $W$-$t$-$b$ interaction is simply a free parameter corresponding to the
operator, 
\bea
g_{Wtb} \: \bar{t} \gamma^\mu W^+_\mu P_L b + h.c. ,
\label{eq:dim4}
\eea
whose value is in principle 
predicted by the underlying strongly coupled theory.  For
extensions which realize the EW symmetry linearly, the coupling can be modified
by dimension six operators \cite{Buchmuller:1985jz}, 
an example of which is,
\bea
\frac{1}{\Lambda^2} \left(  \bar{Q}_3 D_\mu H \right) \gamma^\mu
\left( H^\dagger Q_3 \right) ,
\label{eq:dim6}
\eea
where the parentheses denote contractions of $SU(2)$ indices and
$\Lambda$ represents the scale of new physics effects.
Clearly, replacing the Higgs by its vacuum expectation value reduces
Eq.~(\ref{eq:dim6}) to terms including Eq.~(\ref{eq:dim4}) with
$g_{Wtb} \sim v^2 / \Lambda^2$.  In our analysis,
we focus on the lowest order (dimension 4) effects of Eq.~(\ref{eq:dim4}).
These are likely to be the first effects manifest in precision measurements.
It is worth noting that as one sees from the generic operator above,
most models will also modify other
properties of the top quark, including its couplings to the $Z$ boson and the
Higgs.  In such cases, a precise measurement of $W$-$t$-$b$ can be combined
with other measurements to help unravel the nature of the underlying UV 
physics.  For simplicity, we restrict ourselves to modifications of the
left-chiral interaction, though right-chiral modifications are
straight-forward to include in our analysis.

Since top decays are ineffectual in measuring the $W$-$t$-$b$ interaction
strength, one should use electroweak production processes whose rates are 
directly proportional to the coupling.  Single top production at the
Tevatron and LHC will fill this role.  Discovery at the Tevatron seems
likely by the end of its lifetime, and observation seems certain at the LHC.
The uncertainties in rate at the LHC will be entirely dominated by systematics,
and it is expected that a measurement of $\delta V_{tb} \sim 10\%$ is 
possible \cite{Beneke:2000hk}.  Given the SM prediction inferred from
unitarity at the better than percent level, it is important to consider other
processes from which information about $W$-$t$-$b$ can be extracted.  In
particular, the International Linear Collider
(ILC) represents a clean environment and is expected to achieve
energies on the order of 500 GeV.  It is an ideal place to learn about the
top quark.

However, unlike many other top measurements, a direct test of the $W$-$t$-$b$
coupling is challenging at a $\sim$ 500 GeV $e^+ e^-$ collider.  A scan over
the $t\bar{t}$ threshold region is expected to yield precise
measurements of many top parameters in the SM, including the top mass,
width, and Yukawa coupling (see \cite{Martinez:2002st,Juste:2006sv}
for projections), while above-threshold measurements may constrain
anomalous, non-SM Lorentz structures \cite{Boos:1999ca}.
Nevertheless, only an indirect measurement of the left-handed
$W$-$t$-$b$ coupling is offered from the $t\bar{t}$ threshold region,
by inferring its value from the SM relation and a precise value of the
top width. If, for example, there is a small non-standard decay mode
of top, it will alter the width and distort the inferred coupling.  For
example, if there is a small non-standard decay mode, the top's total width
becomes,
\bea
\Gamma_t & = & \left( \frac{g_{Wtb}}{g} \right)^2 
\frac{G_F m_t^3}{\sqrt{2} 8 \pi} 
\left( 1 - \frac{M_W^2}{m_t^2} \right)^2
\left( 1 + 2 \frac{M_W^2}{m_t^2} \right)
+ \Gamma_{new} ,
\eea
where $\Gamma_{new}$ represents some non-standard decay mode and we
have neglected corrections of order $(m_b/m_t)^2$.  It may be
that this new decay mode can be observed in its own right, but if it is small
or difficult to detect, it may be over-looked.  In that case, the measurement
of $g_{Wtb}$ is actually distorted by the presence of the new physics itself.

Thus, it
would be more desirable to have a direct measurement of $W$-$t$-$b$,
by making use of a process which is directly proportional to it. Close
to the $t\bar{t}$ threshold, sensitivity to the coupling is quite
weak, because the rate is essentially the $t\bar{t}$ production cross
section times the branching ratios for $t \rightarrow W b$, which as explained
above is not sensitive to $g_{Wtb}$.  A $e \gamma$ collider can measure the
rate of single-top production above threshold \cite{Boos:2001sj} to extract
$g_{Wtb}$, but it is also worthwhile to see if the $e^+ e^-$ running mode
can be used to extract useful information below the $t\bar{t}$ threshold.

This article is outlined as follows.  In Section \ref{sec:basics} we
explore $t \bar{t}$ below threshold and demonstrate the underlying dependence
on $g_{Wtb}$.  In Section~\ref{sec:analysis} we examine how the signal may
be extracted from the backgrounds, and determine our sensitivity to
$g_{Wtb}$.  We conclude in Section~\ref{sec:conclusions}.

\section{Leverage from Below the $t \bar{t}$ Threshold}
\label{sec:basics}
\begin{figure}[t]
\begin{center}
\begin{picture}(210,200)(0,0)
\SetColor{Black}
\ArrowLine(0,50)(50,100)
\Text(-2,50)[r]{$e^-$}
\ArrowLine(50,100)(0,150)
\Text(-2,150)[r]{$e^+$}
\Photon(50,100)(121,100){4}{4}
\Text(85,105)[b]{$\gamma , \ Z$}
\SetColor{Blue}
\ArrowLine(121,100)(171,150)
\Text(145,127)[rb]{$t$}
\ArrowLine(171,50)(121,100)
\Text(145,73)[rt]{$t$}
\SetColor{Black}
\Photon(171,150)(211,190){4}{4}
\Text(213,190)[l]{$W^+$}
\ArrowLine(171,150)(211,110)
\Text(213,110)[l]{$b$}
\Photon(171,50)(211,90){4}{4}
\Text(213,90)[l]{$W^-$}
\ArrowLine(211,10)(171,50)
\Text(213,10)[l]{$\bar{b}$}
\end{picture}
\caption{Main contribution to $t\bar{t}$ production at the ILC.}
\end{center}
\label{fig:diag}  
\end{figure}
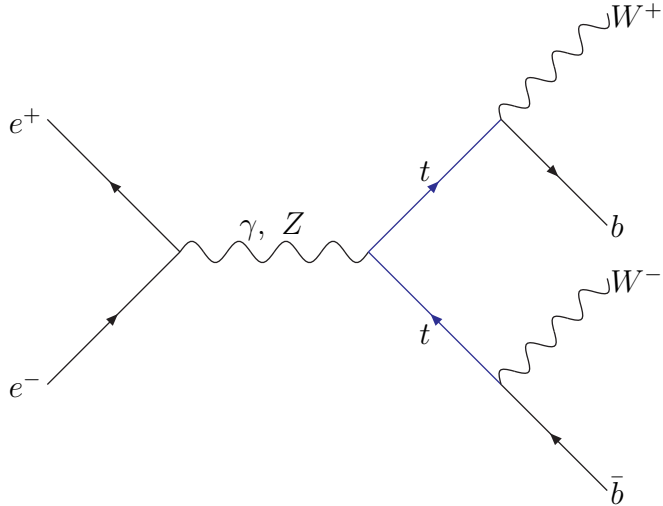

Above the $t\bar{t}$ threshold, final state $W^+  W^-  b  \bar{b}$
events have an uncut cross section of order $500$ fb---
dominated by $t\bar{t}$'s produced from s-channel $\gamma$'s and $Z$'s (see Figure \ref{fig:diag}).
In the narrow top width approximation, the dominant 
piece of the 
cross section factors into a production cross section times the appropriate 
branching ratios:
\be 
\sigma = \sum_{s_1 s_2} \sigma\left( e^+ \ e^-
\rightarrow t(s_1) \bar{t}(s_2) \right) BR \left(t(s_1)
\rightarrow W^+ b \right) BR \left(\bar{t}(s_2) \rightarrow
W^- \bar{b} \right) , \nonumber 
\ee 
where $s_{1,2}$ label the spin state of the $t$ and $\bar{t}$.
Above threshold $\sigma$ has little
or no dependence on $g_{Wtb}$: The $t\bar{t}$ production cross
section is independent of $g_{Wtb}$, and the branching ratios are 
almost exactly unity. 

Just below the $t\bar{t}$ threshold, the intermediate $ t \bar{t}$
diagrams still contribute, along with other non-resonant Feynman
diagrams, to the $W^+ W^- b \bar{b}$ final state. At center-of-mass
energies below $2 m_t$ but still above $m_t$, the total rate is
dominated by contributions from the virtual $t \bar t$ diagrams in a
kinematic configuration where one top is on-shell and the other is
off-shell.  The rate becomes very sensitive to the $W$-$t$-$b$
interaction, essentially because the narrow width approximation is no
longer valid when the top momentum is off-shell. The leading piece in
the narrow width approximation for the virtual top,
\be 
\frac{1}{\left(q_{t*}^2-m_t^2 \right)^2 + m_t^2 \Gamma_t^2} 
\simeq \frac{\pi}{m_t \Gamma_t} \delta\left(q_{t*}^2-m^2\right) ,
\ee 
is zero, and one can no
longer simply disentangle the cross section into production and decay
rates.

\begin{figure}[t]  
\centerline{\includegraphics[width=5in]{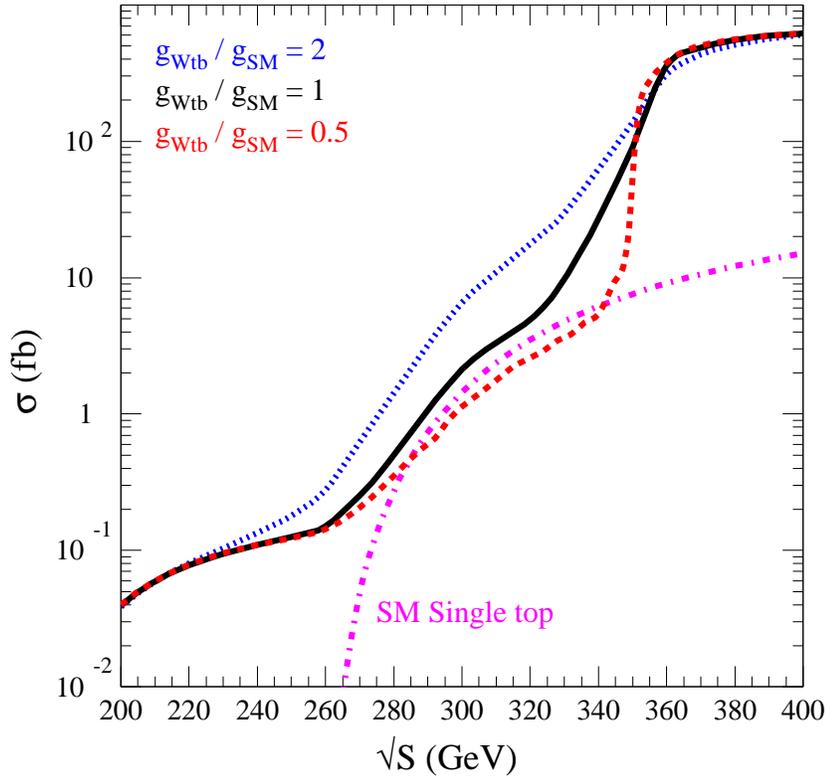}}
\caption{Rates 
for $e^+ \ e^- \rightarrow W^+ b W^- \bar{b}$ as a function
of the center-of-mass energy for $g_{Wtb} = g_{SM}$ (black solid), 
$g_{Wtb} = 2 g_{SM}$ (blue dashed), and $g_{Wtb} = g_{SM} /2 $ (red dotted).
Also shown for reference is the SM single top rate, 
$e^+ e^- \rightarrow t W b$ (violet dash-dot).}
\label{fig:wtb1}  
\end{figure}  

  This is illustrated in Figure~\ref{fig:wtb1}, which plots the cross
section as a function of energy for several values of $g_{Wtb}$,
assuming a $175$ GeV top mass and a $115$ GeV Higgs mass.  All
analysis was performed using the MadEvent package
\cite{Maltoni:2002qb} at tree level.  The cross-sections asymptote to the same
value at both ends of the energy spectrum, as on-shell $t\bar{t}$
production dominates close to threshold and graphs not involving top
dominate far below threshold. Both of these extremes are independent
of the $W$-$t$-$b$ coupling, while energies in between these two
extremes are suitable to measure $g_{Wtb}$. The inflection points in
the intermediate region are due to the turn on of single-top and
associated $W$ production (through graphs that do not contain a virtual top)
at their 255 GeV threshold, and large $t
t^{\star}$ contributions that dominate near
$\sim 350$ GeV. 

For our analysis, we assume a relatively large luminosity 100
fb$^{-1}$ of data collected at a single center of mass energy. We
avoid the region very close to $2 m_t$ (despite its large rate),
because the details of the transition from off-shell to on-shell do
depend sensitively on the top width, which could obscure $g_{Wtb}$ if
there are non-standard decay modes of the top.  Instead, we focus on
the energy $\sqrt{s}= 340$ GeV, just a few GeV above the peak
statistical sensitivity to large deviations in $\delta g_{Wtb}$. Beam
spread effects will not erase our sensitivity, but will contribute
with a worse ratio of signal to background.  The beam energy spread for
a TESLA-like machine is expected to be at the $.1\%$ level
\cite{Thomson:2004ah} and its effect could be compensated by a more highly
optimized choice of $\sqrt{s}$. 

At 340 GeV, the fully interfering electroweak $W^+  W^-  b  \bar{b}$
signal contains contributions from diagrams without any top
propagators, that are reduced by requiring
that the invariant mass of one $W  b$ system reconstruct to a top
mass.   The inclusion of the $t \bar{t}$ 
diagrams enhances the cross section that passes an invariant
top-mass cut by about a factor of two. This is not to say that single-top 
diagrams are unimportant; in fact, true single-top diagrams have maximum 
sensitivity to $g_{Wtb}$, while $t \bar{t}$ diagrams have a sensitivity 
that decreases as $t \bar{t}$ threshold is approached. From 
Figure~\ref{fig:wtb1}, one can see that this sensitivity increases above 
the single top threshold, indicating that the $t \bar{t}$ 
contribution is also important. 

\section{ILC Event Rates, Efficiencies and Backgrounds}
\label{sec:analysis}

We focus on the semi-leptonic six-body final state where one of the W's
decays to a pair of jets and the other $W$ decays into an readily
tagged lepton: $e$, $\mu$ or $\tau$.  To determine the variation of
the signal that passes our cuts as a function of $g_{Wtb}$ and
$\Gamma_t$, we calculate the variation in the rate of the
intermediate final state
$W^+ b W^- \bar{b}$ and multiply by an overall efficiency
factor. The efficiency is the Standard Model ratio of the
six-body final state divided by the four-body final state, and 
intuitively is just the product of branching fractions and 
$b$-tagging efficiencies. 
 
This efficiency factor oversimplifies the situation, as there can be
interference in the full six-body final state, as well as new
contributions from diagrams without the assumed $W^+ W^- b \bar{b}$
intermediate state that manage to pass our cuts.  These separate
effects may change the dependence on $g_{Wtb}$ and $\Gamma_t$, but for
small shifts of $g_{Wtb}$ are negligible and subdominant compared to
our predicted statistical uncertainties, and less important compared
to higher-order perturbative corrections.  For example, shifting the
top width by $100$~MeV alters our efficiency from $14.5 \%$ to $14.4
\%$.  The purity of our sample (the ratio formed from the cross section of the fully interfering six-body final state to the decayed, on-shell $W^+ W^- b \bar{b}$ state)  is of order $\sim 90 \%$. Thus, while the full
six-body simulation will ultimately be important to extract the
correct value of $g_{Wtb}$, we do not expect that our use of the
four-body simulation leads to large changes in our estimation of the
sensitivity with which $g_{Wtb}$ can be extracted. 

In the four-body final state, we impose $|y| <2$ and $p_t > 10$ GeV
cuts on the $b$ jets and require that a single top mass is
reconstructed to within 10 GeV, without assuming charge identification
of the $b$ jets.  In the six body final state with a negative sign
lepton, we model the detector acceptance by requiring
$p_t > 10$ GeV and $|y| < 2$ on all visible final state
particles. In order to remove the non-top initiated backgrounds, we
demand that both the untagged jets and the lepton/missing-energy 
system have an invariant mass within $\pm 5$ GeV of the $W$ mass, 
and that one top can be reconstructed from one of
these $W$'s with either sign $b$ jet. The widths of the invariant mass
acceptances are sufficiently broad for the few GeV jet energy
uncertainty expected at the ILC \cite{Thomson:2004ah}. 
We also assume a particle flow analysis that eliminates the need for
strong lepton/jet isolation.  We assume a $b$-tagging efficiency of
$70 \%$, and find that the SM is expected to produce
 $\sim 220$ events passing the cuts for 100 fb$^{-1}$ of luminosity.

\begin{figure}[t]  
\centerline{\includegraphics[width=4in]{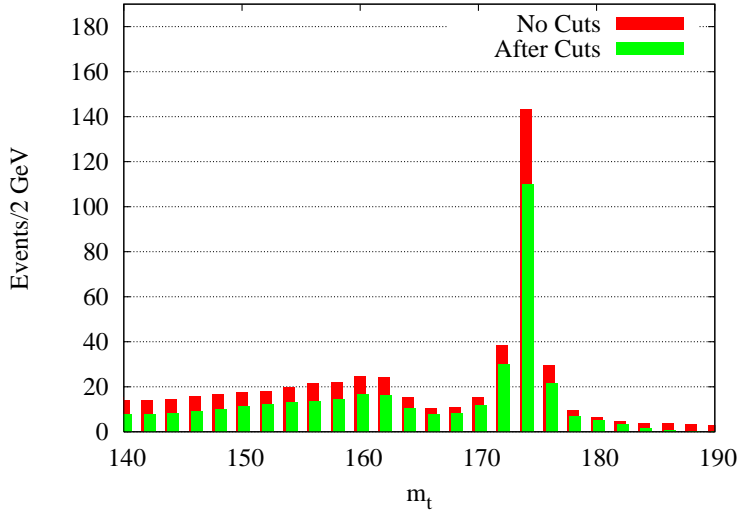}}
\caption{The invariant mass of the hadronic ``top'' mass. 
The ``No Cuts'' histogram shows the distribution of 
the $jjb$ invariant mass, while 
the ``After Cuts'' histogram shows the invariant mass 
of the $jjb$ system after both $W$'s and a single top have been 
reconstructed.}
\label{fig:cuts}  
\end{figure}

The dominant background that is independent of the $W$-$t$-$b$
coupling comes from diagrams with an intermediate Higgs or $Z$ 
that decays to $b \bar{b}$, and could be
eliminated by subtracting events with $b \bar{b}$ that have an
invariant mass close to the (assumed known) Higgs and $Z$ masses.
Still, most of these events are already cut out by demanding a single
$t$ mass reconstruction (either leptonic or hadronic) as shown 
in Figure~\ref{fig:cuts}, which shows the invariant mass of the 
reconstructed hadronic ``top'', which is the 
invariant mass of the $jjb$ system. The bump near 160 GeV represents the 
contribution from $t \bar{t}$ diagrams where the leptonic top is on-shell. 
We have considered the background from $4j + l + \nu$, where two of the
light jets are mistagged as $b$'s.  We find that the number
of events that pass our cuts {\em before} applying the mis-tag probability
is slightly smaller than the number of expected signal events,
and thus the fake contribution after applying the double-mis-tagging 
probability is negligible.

\begin{figure}[t]  
\centerline{\includegraphics[width=4in]{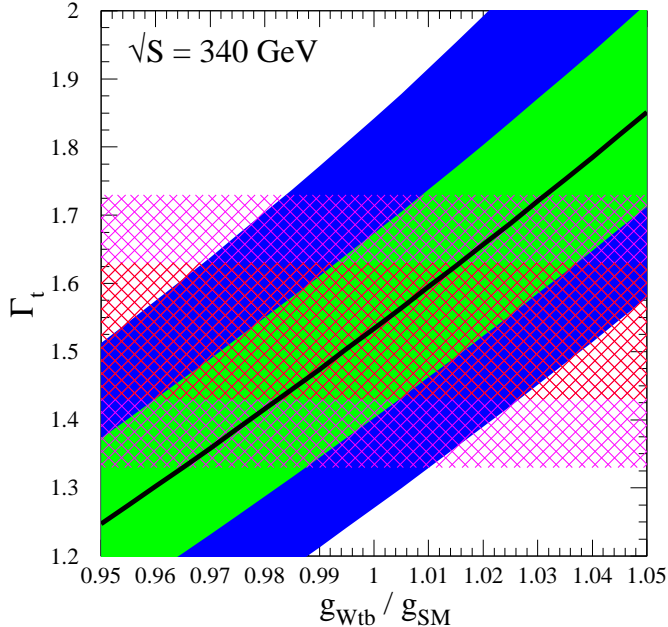}}
\caption{Curve corresponding to the region of the plane
of  $g_{Wtb}$ and $\Gamma_t$ which is degenerate with the SM event
rate and its 1$\sigma$ and 2$\sigma$ deviations as the solid bands.  
Also overlaid is an
expected measurement of $\Gamma_t$ from the on-shell threshold scan with
an uncertainty of $100$ MeV as the cross-hatched bands.} 
\label{fig:wtb2}  
\end{figure}

The number of events observed will depend strongly on $g_{Wtb}$,
the top mass, the top width, and (to some extent) the Higgs mass.
It is expected that the ILC will determine the top and
Higgs masses to order 100 MeV or better, which is enough to render the
uncertainty in the rates from the uncertainty in
these parameters order 1/10th of our 
expected statistical uncertainties, and thus are negligible.  
The remaining dependence on the
width and $g_{Wtb}$ allows us to determine a combination of both these
quantities.  In Figure~\ref{fig:wtb2} we
present the contours of constant event numbers in the plane of
$g_{Wtb}$ and $\Gamma_t$ which reproduce the expected SM event rate of
$\sim 220$ events (as the solid line).  
Also shown as the solid bands are the contours corresponding to
1$\sigma$ and 2$\sigma$ deviations from such a measurement (assuming
that the SM rate is observed and considering purely statistical
uncertainties since we expect these to dominate).  
The result is the expected bound one would obtain on
$g_{Wtb}$ and $\Gamma_t$ if the SM rate is observed. 
Combining the below threshold cross section measurement with the $\Gamma_t$
extracted from the above-threshold scan allows us to extract both
$g_{Wtb}$ and $\Gamma_t$ independently. 
Alternately, one can go to lower energies where the sensitivity to
$\Gamma_t$ is less, though at the price of the loss of some
statistics.  
We have made the conservative assumption that $\Gamma_t$ will be known to
order $\pm 100$ MeV from the above threshold scan \cite{Juste:2006sv}.
From Figure~\ref{fig:wtb2}, we see that given this assumption, $g_{Wtb}$ 
can be measured to the $ 3\%$ level, which would represent better than a
factor of  2 improvement compared to the LHC, and a major improvement
in our understanding of the $W$-$t$-$b$ interaction.

\section{Conclusions} 
\label{sec:conclusions}

The mass of the top quark is a strong indication that the top may play
a fundamental role in the mechanics behind EWSB, or, if not, magnify
the effects of any new physics through the lens of the large top
Yukawa. If this new physics is sufficiently decoupled, shifts in the
SM-like top couplings may be the only evidence left behind; it is no
surprise that measuring the properties of the top quark will remain a
collider focus for the next few decades.

\begin{figure}[t]  
\centerline{\includegraphics[width=4in]{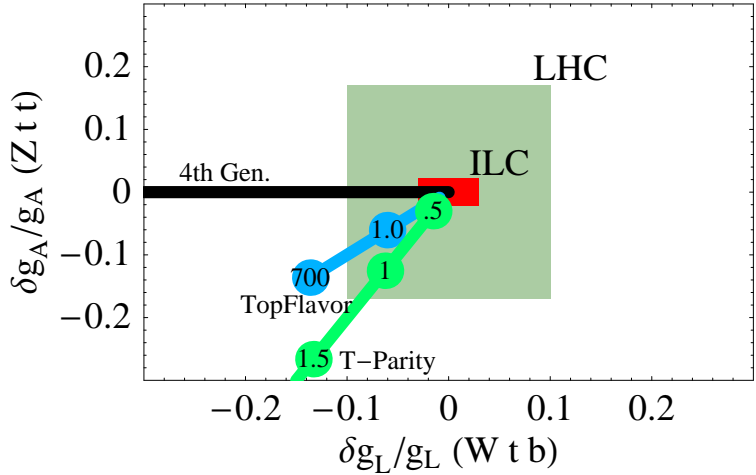}}
\caption{Expected bounds on SM-like couplings, axial $Z$-$t$-$\bar{t}$
and left-handed $W$-$t$-$b$, from direct LHC and ILC measurements. LHC
bounds are shown in olive, ILC bounds in red. Superimposed are
predicted deviations from representative models described in the text.}
\label{fig:limits}  
\end{figure}

Although single top production is usually ignored in $e^+ e^-$
collisions, a measurement of $g_{Wtb}$ is not out of reach at the
ILC. A significant amount of leverage is, counter-intuitively,
provided by $t \bar{t}$ production {\it below} threshold. The results
of our analysis are shown in Figure~\ref{fig:limits}: we compute a
1$\sigma$ error in the $W$-$t$-$b$ coupling of order a few
percent. This constraint is on par with the indirect bound on
$g_{Wtb}$ coming from the threshold measurement of the top-width,
though the direct bound we present does not depend on a detailed
understanding of all top decay modes and branching fractions, and thus
is complementary to the measurement of the top width.

Figure~\ref{fig:limits} shows the new expected bounds on the
SM-like top axial $Z$-$t$-$\bar{t}$ and left-handed $W$-$t$-$b$ interactions
and the discriminating power the new bounds can place on new physics
models. We include our results with the 1$\sigma$ constraints on the
independently varied axial $Z$-$t$-$\bar{t}$ coupling from the LHC
\cite{Baur:2005wi} and ILC \cite{Abe:2001nq}, and the direct
constraints on the left-handed $W$-$t$-$b$ coupling from the LHC
\cite{Beneke:2000hk}.  Predicted deviations from a few representative
models are also superimposed: a Little Higgs model with T-parity, a
model of topflavor, and a model with a sequential fourth generation whose
quarks mix substantially with the third family. The Little
Higgs $T$-parity model has a heavy top-partner, $T$, with mass 500 GeV
(the numbers on the plot indicate the strength of the $h$-$T$-$t$ interaction)
\cite{Berger:2005ht}; the topflavor model has a mixing angle
$\sin{\phi} = .9$ (numbers indicate the mass of the heavy $Z^\prime$)
\cite{Chivukula:1995gu}.  
Top-seesaw models have the same mixing effect as the
Little Higgs model, and thus trace out the same line in the plane of
deviations in the $Z$-$t$-$\bar{t}$ and $W$-$t$-$b$ as the seesaw model
parameters are varied.

Many improvements on our approximate results are possible. In particular,
higher order QCD and EW corrections to the signal will be essential to
include in a realistic analysis in order to obtain the desired
accuracy in $g_{Wtb}$, particularly effects from initial state
radiation and beamstrahlung, which will likely require stronger
$\theta$ dependent cuts to cut down the resultant
backgrounds. However, a consideration of the $4 j b \bar{b}$ final
state could add a comparable amount of statistics to the semileptonic
sample we've considered. 
Further, one could avoid the need for a large set of data at a single
energy by performing several measurements of smaller integrated luminosities
at a range of energies.  This could allow one to use the energy dependent 
shape of the cross section as well as the normalization, and could potentially
allow one to achieve comparable accuracy with much less integrated luminosity.
We leave such refinements for future work.

\vspace*{1cm}

\noindent
{\bf \large Acknowledgements}\\[0.2cm]
It is a pleasure to acknowledge conversations with A. Juste, T. LeCompte,
S. Magill, F. Petriello, Z. Sullivan, and especially F. Maltoni.
Work at ANL is supported in part by the US DOE, Division of HEP, 
Contract W-31-109-ENG-38.

\end{document}